\documentclass[reprint,superscriptaddress,longbibliography,prl]{revtex4-1}

\usepackage{amsmath}
\usepackage{amssymb}
\usepackage{graphicx}
\usepackage{color}
\usepackage{bm}         % bold math
\usepackage{MnSymbol}

\newcommand{\mbf}[1]{\bm{#1}}
\newcommand{\cS}{{\cal S}}
\newcommand{\avg}[1]{\langle #1 \rangle}
\newcommand{\cg}{\textnormal{\textsl{g}}}
\newcommand{\vG}{{\check G}}
\newcommand{\aCom}[2]{\{ #1 , #2 \}}

\begin{document}
    
%\preprint{APS/123-QED}

\title{Continuous-Variable Entanglement Test in Driven Quantum 
Contacts}

\author{Hongxin Zhan}
\affiliation{Fachbereich Physik, Universit\"{a}t Konstanz, 
    D-78457 Konstanz, Germany}

\author{Mihajlo Vanevi\' c}
\affiliation{Department of Physics, University of Belgrade, 
    11158 Belgrade, Serbia}

\author{Wolfgang Belzig}
\affiliation{Fachbereich Physik, Universit\"{a}t Konstanz, 
    D-78457 Konstanz, Germany}

\date{\today}% It is always \today

%%%%%%%%%%%%%%%%%%%%%%%%%%%%%%%%%%%%%%%%%%%%%%%%%%%%%%%%%%%%%%%%%%%%%%%%
\begin{abstract}
The standard entanglement test using the Clauser-Horne-Shimony-Holt
inequality is known to fail in mesoscopic junctions at finite
temperatures. Since this is due to the bidirectional particle flow, a
similar failure is expected to occur in an ac-driven contact. We develop a continuous-variable entanglement test suitable for electrons
and holes that are created by the ac drive. At low enough temperatures the generalized Bell inequality is violated in junctions with low
conductance or small number of transport channels and with ac voltages
which create few electron-hole pairs per cycle. Our ac-entanglement
test depends on the total number of electron-hole pairs  
and on the distribution of probabilities of pair creations similar to 
the Fano factor.
\end{abstract}
%%%%%%%%%%%%%%%%%%%%%%%%%%%%%%%%%%%%%%%%%%%%%%%%%%%%%%%%%%%%%%%%%%%%%%%%

\pacs{Valid PACS appear here}


\maketitle

%%%%%%%%%%%%%%%%%%%%%%%%%%%%%%%%%%%%%%%%%%%%%%%%%%%%%%%%%%%%%%%%%%%%%%%%
%%%%%%%%%%%%%%%%%%%%%%%%%%%%%%%%%%%%%%%%%%%%%%%%%%%%%%%%%%%%%%%%%%%%%%%%
%%%%%%%%%%%%%%%%%%%%%%%%%%%%%%%%%%%%%%%%%%%%%%%%%%%%%%%%%%%%%%%%%%%%%%%%
%%%%%%%%%%%%%%%%%%%%%%%%%%%%%%%%%%%%%%%%%%%%%%%%%%%%%%%%%%%%%%%%%%%%%%%%
Quantum mechanics in fact is a nonlocal theory. In 1935 Einstein, 
Podolsky, and Rosen put forward the EPR paradox 
\cite{einstein_can_1935} 
to question the completeness of quantum mechanics which disobeys, 
in their view apparent, the principle of local realism. 
They argued that quantum mechanics should be completed with some 
hidden variables.
In 1964 Bell derived an inequality \cite{bell_on-EPR_1964} to actually 
test the principle of local realism. Remarkably, he found that in some 
cases the physical results based on the principle of local realism are 
inconsistent with the predictions of quantum mechanics. 
In later years, the experiments on quantum entanglement  
\cite{freedman_experimental_1972,*aspect_experimental_1981,%
  *aspect_experimental_1982,*rarity_experimental_1990,%
  *rowe_experimental_2001,*weihs_violation_1998}
proved the existence of nonlocality, including the recent 
loophole-free or human-choice-driven tests \cite{hensen_loophole-free_2015,%
  giustina_significant-loophole-free_2015,shalm_strong_2015,TheBIGBellTestCollaboration:2018ge}.
Besides their fundamental importance, quantum entanglement and 
nonlocality have attracted a lot of attention over the past two 
decades in the context of quantum computation 
\cite{steane_quantum_1998}, teleportation 
\cite{bennett_teleporting_1993,boschi_experimental_1998}, and cryptography 
\cite{Gisin-RevModPhys74-02}.
While most of the tests are performed using quantum 
optics, it still remains a challenge to detect entanglement 
within the Fermi sea in mesoscopic systems
\cite{Burkard:2000un,loss_probing_2000,Recher:2001je,bena_quantum_2002,%
	samuelsson_orbital_2003,samuelsson_two_2004,beenakker_proposal_2003,%
	*beenakker_optimal_2005,Samuelsson:2009kh,klich_quantum_2009}.
Some of these concepts have even been tested experimentally
\cite{Hofstetter2009,Gabor2009,Herrmann:2010dm,Hofstetter:2011hg,%
	Simmons2011,Das:2012gp,Schindele:2012wg,Fulop:2015te,Gramich:2016wc}.

The most common way to detect the entanglement is to test for the 
violation of the Clauser-Horne-Shimony-Holt (CHSH) inequality
\cite{clauser_proposed_1969}. 
%
%%%%%%%%%%%%%%%%%%%%%%%%%%%%%%%%%%%%%%%%%%%%%%%%%%%%%%%%%%%%%%%%%%%%%%%%
%%%%%%%%%%%%%%%%%%%%%%%%%%%%%%%%%%%%%%%%%%%%%%%%%%%%%%%%%%%%%%%%%%%%%%%%
%%%%%%%%%%%%%%%%%%%%%%%%%%%%%%%%%%%%%%%%%%%%%%%%%%%%%%%%%%%%%%%%%%%%%%%%
\begin{figure}[t]
    %\centering
    \includegraphics[scale=0.5]{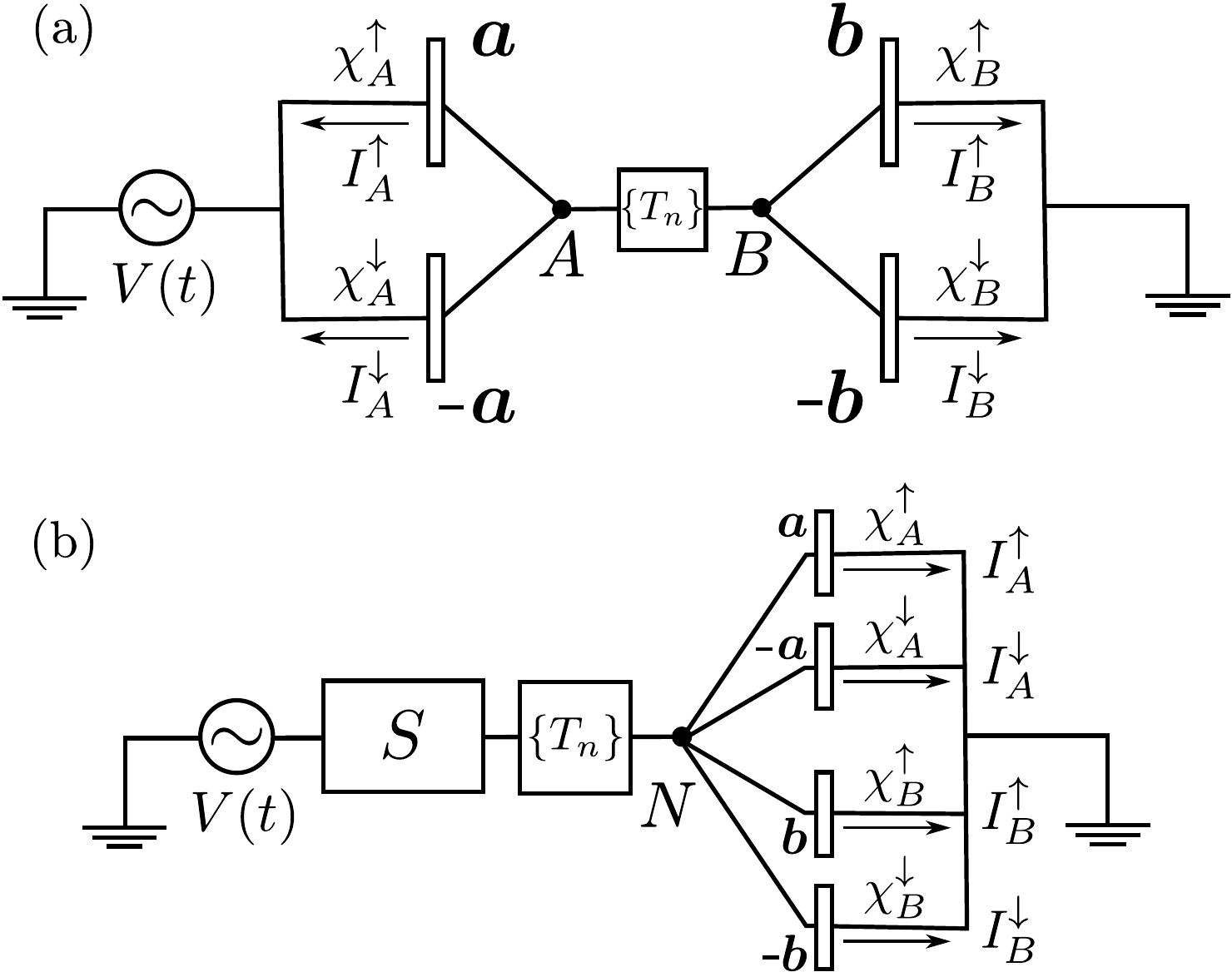}
    \caption{\label{fig:setups}Setups for the generalized ac Bell test: 
        (a) normal junction with transmission eigenvalues $\{T_n\}$ and 
        (b) superconductor -- normal-metal beam splitter. 
        In both cases, the
        test observables are the differences between the numbers of 
        spin-up
        and spin-down particles along directions $\pm\mbf a$ and 
        $\pm\mbf b$
        detected in the leads \textit{A} (Alice) and \textit{B} (Bob).}
\end{figure}
%%%%%%%%%%%%%%%%%%%%%%%%%%%%%%%%%%%%%%%%%%%%%%%%%%%%%%%%%%%%%%%%%%%%%%%%
%
In mesoscopic junctions as shown in Fig.~\ref{fig:setups}, the CHSH inequality reads
%%%%%%%%%%%%%%%%%%%%%%%%%%%%%%%%%%%%%%%%%%%%%%%%%%%%%%%%%%%%%%%%%%%%%%%%
%\begin{equation}\label{BI0}
$|C(\mbf a,\mbf b) + C(\mbf a,\mbf b') + 
C(\mbf a',\mbf b) - C(\mbf a',\mbf b')| \le 2$
%\end{equation}
%%%%%%%%%%%%%%%%%%%%%%%%%%%%%%%%%%%%%%%%%%%%%%%%%%%%%%%%%%%%%%%%%%%%%%%%
where 
%%%%%%%%%%%%%%%%%%%%%%%%%%%%%%%%%%%%%%%%%%%%%%%%%%%%%%%%%%%%%%%%%%%%%%%%
%\begin{equation}
$C(\mbf n_A,\mbf n_B) =
\langle (N_A^\uparrow - N_A^\downarrow)
    (N_B^\uparrow - N_B^\downarrow) \rangle
/ \langle (N_A^\uparrow + N_A^\downarrow)
    (N_B^\uparrow + N_B^\downarrow) \rangle$
%\end{equation}
%%%%%%%%%%%%%%%%%%%%%%%%%%%%%%%%%%%%%%%%%%%%%%%%%%%%%%%%%%%%%%%%%%%%%%%%
are the spin correlators and $N_j^\sigma$ are the numbers of 
quasiparticles with spin 
projections $\sigma=\uparrow,\downarrow$ in direction $\mbf n_j$ 
detected in the lead $j=A,B$. Assuming 100\% efficient detectors, the spin correlators can be expressed in 
terms of the average current $I_0$ and current noise power 
$S_0$ \cite{chtchelkatchev_bell_2002}, 
$C(\mbf n_A,\mbf n_B)=-\mbf n_A\cdot\mbf n_B S_0 / (S_0+2\tau I_0^2)$,
where $\tau$ is the measurement time. 
What is peculiar about this result is that it reduces to 
$C(\mbf n_A,\mbf n_B)=-\mbf n_A\cdot \mbf n_B$ 
in an ac driven system with no dc bias ($I_0=0$).
In that case, the CHSH inequality can always be 
maximally violated with the left-hand side equal to $2\sqrt{2}$ 
if the angles 
between the spin polarization directions in the leads are chosen as 
$\theta_{\mbf a\mbf b} = \theta_{\mbf a'\mbf b} = 
\theta_{\mbf a\mbf b'}=\pi/4$ and $\theta_{\mbf a'\mbf b'}=3\pi/4$.
However, this violation does not mean entanglement because it cannot
single out the entangled pair if there are many entangled pairs.
In addition, it is implausible that the violation is independent of 
the amplitude and the frequency of the ac voltage and even the shape of 
the drive. The reason why the CHSH inequality fails to reveal the 
entanglement is because it requires unidirectional particle
flow in the leads, a condition which is not satisfied 
when we apply ac 
voltage to the system. Indeed, ac drive leads to fluctuations with 
$|N_j^\uparrow - N_j^\downarrow| > |N_j^\uparrow + N_j^\downarrow|$
which are forbidden in the Bell test \cite{chtchelkatchev_bell_2002}.
The same problem occurs in a dc biased junction at finite temperatures
\cite{hannes_finite-temperature_2008}. To detect entanglement in an ac 
driven junction, a generalized Bell test is needed which is free
from this restriction.

%%%%%%%%%%%%%%%%%%%%%%%%%%%%%%%%%%%%%%%%%%%%%%%%%%%%%%%%%%%%%%%%%%%%%%%%
%%%%%%%%%%%%%%%%%%%%%%%%%%%%%%%%%%%%%%%%%%%%%%%%%%%%%%%%%%%%%%%%%%%%%%%%

\begin{figure}[t]
  %\centering
  \includegraphics[scale=0.7]{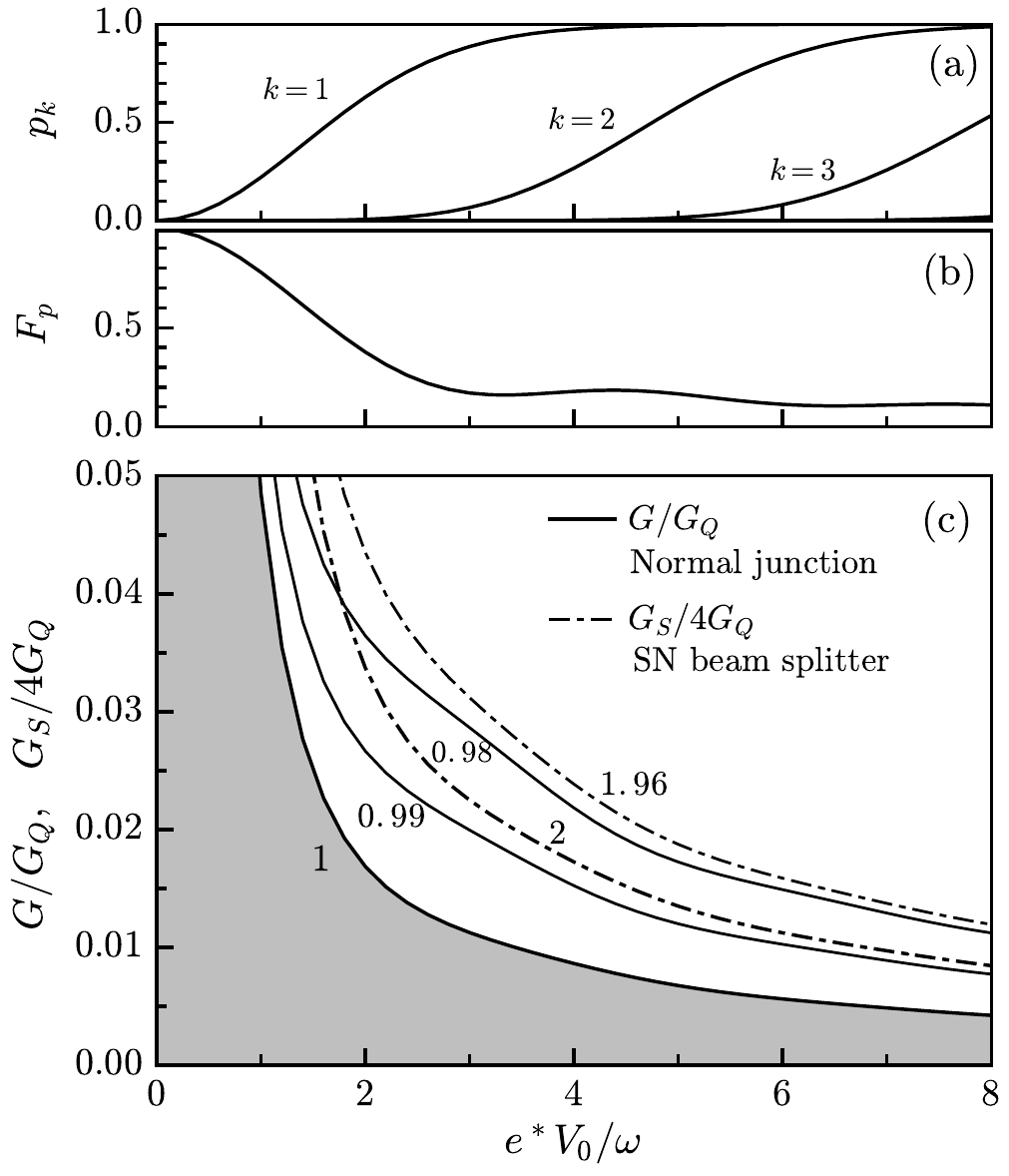}
  \caption{\label{fig:BellTestTunnel}(a) Probabilities of electron-hole 
    pair creations $p_k$ and (b) parameter 
    $F_p=\sum_k p_k(1-p_k)/\sum_k p_k$ 
    as a function of the amplitude $V_0$ for harmonic drive 
    $V(t) = V_0 \cos(\omega t)$. 
    (c) Test of a generalized Bell inequality: Normal junction 
    [solid lines, see 
    Eq.~\eqref{eq:BItunN} and Fig.~\ref{fig:setups}(a)] and \textit{SN} beam 
    splitter geometry [dash-dotted lines, see Eq.~\eqref{eq:BItunSN} 
    and Fig.~\ref{fig:setups}(b)]. 
    The generalized Bell 
    inequality is violated for junction conductances and applied 
    voltages that are below the lines shown in (c). 
    Fano factors $F=1$, $0.99$, $0.98$ ($F_S=2$, $1.96$) for the normal 
    (superconducting) junction are shown in the plot. 
    The effective charge is $e^*=e$ for the normal and $e^*=2e$ for the
    \textit{SN} junction.}
\end{figure}

In this Letter, we study two different setups for a generalized Bell 
test in ac driven systems at low temperatures. The setup shown in
Fig.~\ref{fig:setups}(a) is a normal junction with transmission
eigenvalues $\{T_n\}$ biased by the periodic ac voltage $V(t)$ 
with frequency $\omega$ and zero average. The second setup,
shown in Fig.~\ref{fig:setups}(b), is an ac-driven superconductor (\textit{S}) 
-- normal-metal (\textit{N}) beam
splitter which emits entangled pairs of electrons or holes. 
In both cases, the spin-polarized 
particles along directions $\pm\mbf a$ and $\pm\mbf b$ are detected in 
the leads \textit{A} (Alice) and \textit{B} (Bob). The test observables are the 
differences between the numbers of spin-up and spin-down particles  
$\hat A \equiv \hat N_A^\uparrow - \hat N_A^\downarrow 
= \int dt(\hat I_A^\uparrow - \hat I_A^\downarrow)/e$ and 
$\hat B \equiv \hat N_B^\uparrow - \hat N_B^\downarrow 
= \int dt(\hat I_B^\uparrow - \hat I_B^\downarrow)/e$. 
% detected by Alice and Bob. 
%We consider the periodic ac bias voltage $V(t)$ with 
%the frequency $\omega$ and zero average. 
In contrast to the 
dc bias which at zero temperature leads to unidirectional electron  
transport \cite{bednorz_proposal_2011,lorenzo_full_2005},  
the ac drive generates electron-hole pairs in the junction
\cite{vanevic_elementary_2007,*vanevic_elementary_2008,%
vanevic_elementary_2016,vanevic_control_2012,vanevic_electron_2016}. 
The particle flow in the leads is therefore bidirectional which renders 
the standard Bell test inapplicable. 
In this Letter, 
we derive a generalized Bell inequality for the proposed setups, 
%%%%%%%%%%%%%%%%%%%%%%%%%%%%%%%%%%%%%%%%%%%%%%%%%%%%%%%%%%%%%%%%%%%%%%%%
\begin{equation}\label{eq:BI-setup12}
\sqrt{3}(2-\sqrt{2})/4 \le \sqrt{ 1 - \avg{A^2}/\avg{A^4} },
\end{equation}
%%%%%%%%%%%%%%%%%%%%%%%%%%%%%%%%%%%%%%%%%%%%%%%%%%%%%%%%%%%%%%%%%%%%%%%%
whose violation implies the presence of entangled particles in the 
system. 
Since Eq.~\eqref{eq:BI-setup12} requires access to the 
fourth-order spin current correlations, we anticipate the use of 
low-Ohmic spin-polarized contacts which convert the spin 
into charge currents or testing the spin-dependent chemical potentials 
using ferromagnetic tunnel junctions 
\cite{Das:2012gp,Arakawa_shot_noise_prl2015}.
Expressing the moments $\avg{A^2}$ and $\avg{A^4}$ in terms of 
transmission eigenvalues and properties of the drive, 
we find 
%%%%%%%%%%%%%%%%%%%%%%%%%%%%%%%%%%%%%%%%%%%%%%%%%%%%%%%%%%%%%%%%%%%%%%%%
\begin{equation}\label{eq:BI-final}
\frac{\sqrt{2}-1}{4} \le 
\sqrt{ \frac{X_i}{1 + 6 X_i} }.
\end{equation}
%%%%%%%%%%%%%%%%%%%%%%%%%%%%%%%%%%%%%%%%%%%%%%%%%%%%%%%%%%%%%%%%%%%%%%%%
Here, 
%%%%%%%%%%%%%%%%%%%%%%%%%%%%%%%%%%%%%%%%%%%%%%%%%%%%%%%%%%%%%%%%%%%%%%% 
$X_i = \sum_{n,k} P_{n,k}^{(i)} - \sum_{n,k} P_{n,k}^{(i)2} 
/ \sum_{n,k} P_{n,k}^{(i)}$
%%%%%%%%%%%%%%%%%%%%%%%%%%%%%%%%%%%%%%%%%%%%%%%%%%%%%%%%%%%%%%%%%%%%%%%%
where $i=1$ ($2$) stands for the normal (superconducting) junction 
shown in Fig.~\ref{fig:setups}. 
The probabilities  $P_{n,k}^{(1)} = p_k T_n R_n$ and 
$P_{n,k}^{(2)} = p_k R_n^A/4$, where $R_n=1-T_n$ and
$R_n^A = T_n^2/(2-T_n)^2$ are normal and Andreev reflection 
coefficients; $p_k$ ($k=1,2,\ldots$) are the
probabilities of electron-hole pair creations which depend on the
details of the drive 
\cite{vanevic_elementary_2007,*vanevic_elementary_2008,vanevic_control_2012}.
For the harmonic drive, the dependence of the $p_k$ on the drive amplitude 
is shown in Fig.~\ref{fig:BellTestTunnel} (a).
 
To derive Eqs.~\eqref{eq:BI-setup12} and~\eqref{eq:BI-final}, we first 
obtain the statistics of $\hat A$ and $\hat B$ in the junctions 
at hand. The statistics is computed by using the circuit theory of 
quantum transport. We assign the counting fields $\chi_A^\sigma$ and 
$\chi_B^\sigma$ ($\sigma = \uparrow,\downarrow$) to the spin-polarized 
leads of Alice and Bob, see Fig.~\ref{fig:setups}. 
Since we are only interested in the
differences between the numbers of spin-up and spin-down particles, 
we can set $\chi_A^\uparrow = -\chi_A^\downarrow = \chi_A$ and 
$\chi_B^\uparrow = -\chi_B^\downarrow = \chi_B$, where 
$\chi_A$ and $\chi_B$ are the counting fields which determine the 
statistics of $\hat A$ and $\hat B$.  
For the cumulant generating functions $\cS_1$ and $\cS_2$ of the normal 
and the superconducting junction shown in Fig.~\ref{fig:setups} we find 
\cite{supplemental}
%%%%%%%%%%%%%%%%%%%%%%%%%%%%%%%%%%%%%%%%%%%%%%%%%%%%%%%%%%%%%%%%%%%%%%%%
\begin{multline}\label{CGF12}
\cS_i(\chi_A,\chi_B) 
= 
M \sum_{n,k} \ln\bigg( 1 + P_{n,k}^{(i)}
\\
\times \sum_{\alpha,\beta = \pm 1} 
\frac{1+\alpha\beta\mbf a \cdot\mbf b}{2} 
(e^{i\alpha\chi_A-i\beta\chi_B}-1) \bigg).
\end{multline}
%%%%%%%%%%%%%%%%%%%%%%%%%%%%%%%%%%%%%%%%%%%%%%%%%%%%%%%%%%%%%%%%%%%%%%%%
%
Here, $M=\tau\omega/\pi$ and we assumed low temperature $T_e\ll\omega$ and 
$\omega \ll \Delta$ in the superconducting case, which is in the experimentally accessible range.
We later discuss the effect of finite temperature along the lines of electron-hole
pair creation in Ref.~\cite{vanevic_elementary_2007}.
In the remainder of the Letter we set $M=1$ which is the optimal
value for the Bell test. Indeed, as the measuring time $\tau$ is increased the
number of particles detected also increases which will destroy the
entanglement test. Experimentally, the inverse measuring time $1/\tau$ is related to the bandwidth of the detection setup. Hence, $M=1$ can be achieved by measuring
the cumulants of $\hat A$ and $\hat B$ in a large bandwidth $1/\tau\sim\omega$, which is equivalent to 
dividing the result by $M$ and justifies the setting of $M=1$ in the following.

Next we apply the generalized Bell inequality  
\cite{bednorz_proposal_2011}
%%%%%%%%%%%%%%%%%%%%%%%%%%%%%%%%%%%%%%%%%%%%%%%%%%%%%%%%%%%%%%%%%%%%%%%%
\begin{multline}\label{BI-cumulant}
|\avg{AB(A^2+B^2)} + \avg{A'B(A'^2+B^2)} + \avg{AB'(A^2+B'^2)} 
\\ 
- \avg{A'B'(A'^2+B'^2)}| 
\le 
\avg{A^4} + \avg{A'^4} + \avg{B^4} + \avg{B'^4} 
\\ 
+ \frac{1}{2} \sideset{}{^*}\sum_{C,D,E}
\sqrt{\sqrt{\avg{C^4}}\sqrt{\avg{D^4}} \avg{(D^2-E^2)^2}}.
\end{multline}
%%%%%%%%%%%%%%%%%%%%%%%%%%%%%%%%%%%%%%%%%%%%%%%%%%%%%%%%%%%%%%%%%%%%%%%%
Here, the summation is taken over $C,D,E \in \{A,A',B,B'\}$ and ${}^*$ 
denotes the restriction $D\ne C$ and $E \ne C, D, D'$. 
Importantly, inequality in Eq.~\eqref{BI-cumulant} does not require 
a quantized measurement or unidirectional particle flow and also does not 
pose restrictions on 
the possible values of the observables $A,A',B,B'$. As with any Bell 
test, the symbol $'$ in Eq.~\eqref{BI-cumulant} denotes two different 
settings of the Alice and Bob measuring apparatus. In this case these 
are the two choices of spin-polarization directions $\mbf a$, $\mbf a'$
for Alice and $\mbf b$, $\mbf b'$ for Bob. 
%
%Note, that the efficiency e.g. due to a low polarization drops out from the inequality (4) due to the homogeneity
%
The cumulants of $\hat A$ and $\hat B$ are obtained by taking 
derivatives of the cumulant generating function in Eq.~\eqref{CGF12} 
with respect to the counting fields. 
From the cumulants we can find the moments which appear in 
Eq.~\eqref{BI-cumulant}. We obtain the following relations which are 
valid for both setups:
$\avg{A^{(\prime)}}=\avg{B^{(\prime)}}=0$, 
$\avg{A^{(\prime)4}}=\avg{B^{(\prime)4}}$,
$\avg{A^{(\prime)3}B^{(\prime)}} = \avg{A^{(\prime)}B^{(\prime)3}} 
= -\mbf a^{(\prime)} \cdot \mbf b^{(\prime)}\avg{A^4}$,
$\avg{A^{(\prime)2}B^{(\prime)2}} 
= (1/3) [1+2(\mbf a^{(\prime)} \cdot \mbf b^{(\prime)})^2] \avg{A^4}  
+ (2/3) [1-(\mbf a^{(\prime)} \cdot \mbf b^{(\prime)})^2] \avg{A^2}$. 
Substituting these relations in Eq.~\eqref{BI-cumulant}, we obtain
%%%%%%%%%%%%%%%%%%%%%%%%%%%%%%%%%%%%%%%%%%%%%%%%%%%%%%%%%%%%%%%%%%%%%%%%
$|\mathcal{B}(\mbf a,\mbf b,\mbf a',\mbf b')| 
\le 2 + (2/\sqrt{3})  
\sqrt{ 1 - \avg{A^2}/\avg{A^4} }
\sum_{\mbf c, \mbf d} \sqrt{1 - (\mbf c \cdot \mbf d)^2}$
%%%%%%%%%%%%%%%%%%%%%%%%%%%%%%%%%%%%%%%%%%%%%%%%%%%%%%%%%%%%%%%%%%%%%%%%
where $\mbf c \in \{\mbf a,\mbf a'\}$, $\mbf d \in \{\mbf b,\mbf b'\}$,
and
%%%%%%%%%%%%%%%%%%%%%%%%%%%%%%%%%%%%%%%%%%%%%%%%%%%%%%%%%%%%%%%%%%%%%%%%
$\mathcal{B}(\mbf a,\mbf b,\mbf a',\mbf b')
=
\mbf a \cdot \mbf b + \mbf a' \cdot \mbf b 
+ \mbf a \cdot \mbf b' - \mbf a' \cdot \mbf b'$. 
%%%%%%%%%%%%%%%%%%%%%%%%%%%%%%%%%%%%%%%%%%%%%%%%%%%%%%%%%%%%%%%%%%%%%%%%
Choosing the angles between spin-polarization directions 
$\theta_{\mbf a\mbf b} = \theta_{\mbf a'\mbf b} = 
\theta_{\mbf a\mbf b'}=\pi/4$ and $\theta_{\mbf a'\mbf b'}=3\pi/4$
such that $|\mathcal{B}|$ is maximal, the generalized Bell inequality 
reduces to Eq.~\eqref{eq:BI-setup12}.
Finally, after computing the moments $\avg{A^2} = \partial^2_{i\chi_A} 
\cS |_{\chi=0}$ and $\avg{A^4} = 3\avg{A^2}^2 + \partial^4_{i\chi_A} 
\cS |_{\chi=0}$ we recover Eq.~\eqref{eq:BI-final}. 

%%%%%%%%%%%%%%%%%%%%%%%%%%%%%%%%%%%%%%%%%%%%%%%%%%%%%%%%%%%%%%%%%%%%%%%%
\begin{figure}[t]
  %\centering
  \vspace*{1.5mm}
  \includegraphics[scale=0.7]{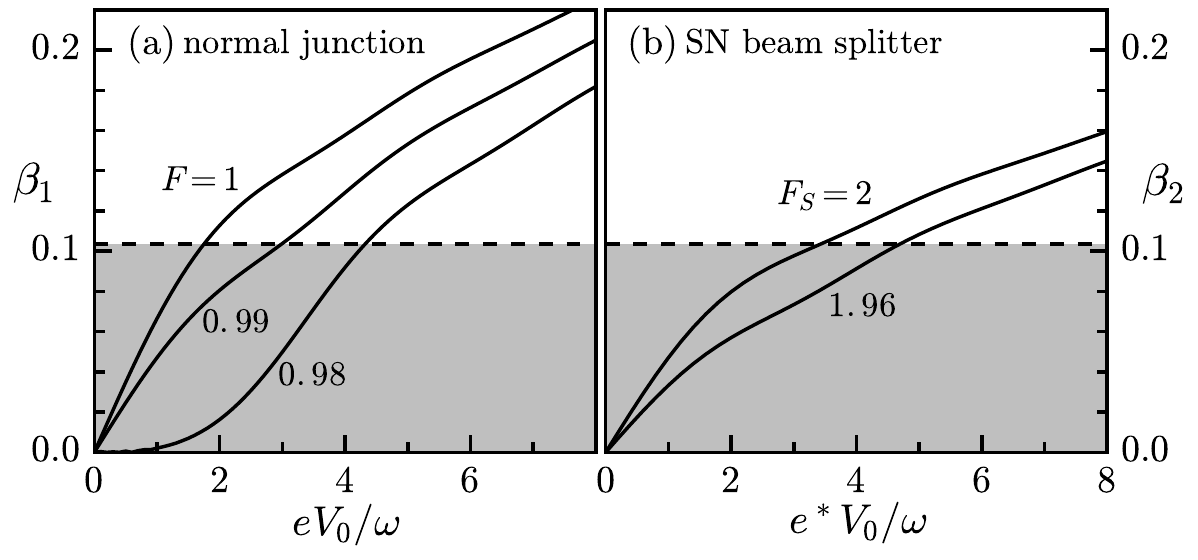}
  \caption{\label{fig:BI-cos-tun-N-SN} 
    Violation of a generalized Bell inequality 
    as a function of the harmonic drive amplitude: (a) normal 
    junction with $G/G_Q = 0.02$ and (b) \textit{SN} beam splitter with 
    $G_S/4G_Q = 0.02$. Inequality is violated in the 
    shaded regions in which $\beta_1$ and $\beta_2$ (solid lines) are 
    smaller than $(\sqrt{2}-1)/4$ (dashed line).}
\end{figure}

%%%%%%%%%%%%%%%%%%%%%%%%%%%%%%%%%%%%%%%%%%%%%%%%%%%%%%%%%%%%%%%%%%%%%%%%
%
Equations~\eqref{eq:BI-setup12} and~\eqref{eq:BI-final} represent a 
generalized Bell test suitable for driven systems 
%at low temperatures 
shown in Fig.~\ref{fig:setups}, where the particle 
flow in the leads is bidirectional due to the presence of electron-hole 
pairs created by the drive. 
When the number of transport channels is large, Eq.~\eqref{eq:BI-final} 
reduces to $(\sqrt{2}-1)/4 \le 1/\sqrt{6}$ which is not violated, as 
one might expect. The same is true when many electron-hole
pairs are created in the system ($p_k=1$ for $k=1,\ldots, N_{eh}$; 
$N_{eh}\gg 1$). To violate Eq.~\eqref{eq:BI-final}, the contribution of 
an entangled pair has to be singled out from the rest. 
This is achieved in a junction with low conductance or small 
number of transport channels and with few particles created per 
voltage cycle.  
In the low conductance limit,  %($R_n \lesssim 1$)
Eq.~\eqref{eq:BI-final} reduces to 
$(\sqrt{2}-1)/4 \le \beta_i$ where to lowest order in $1-F$($1-F_S/2$)
%%%%%%%%%%%%%%%%%%%%%%%%%%%%%%%%%%%%%%%%%%%%%%%%%%%%%%%%%%%%%%%%%%%%%%%%
\begin{equation}\label{eq:BItunN}
\beta_1^2 =  
\frac{G}{G_Q}\sum_k p_k - (1-F_p)(1-F)
\end{equation}
%%%%%%%%%%%%%%%%%%%%%%%%%%%%%%%%%%%%%%%%%%%%%%%%%%%%%%%%%%%%%%%%%%%%%%%%
and 
%%%%%%%%%%%%%%%%%%%%%%%%%%%%%%%%%%%%%%%%%%%%%%%%%%%%%%%%%%%%%%%%%%%%%%%%
\begin{equation}\label{eq:BItunSN}
\beta_2^2 = \frac{1}{4}   
\left(\frac{G_S}{2G_Q}\sum_k p_k - (1-F_p)(1-F_S/2) \right)
\end{equation}
%%%%%%%%%%%%%%%%%%%%%%%%%%%%%%%%%%%%%%%%%%%%%%%%%%%%%%%%%%%%%%%%%%%%%%%%
for the normal junction and the \textit{SN} beam splitter, respectively. 
Here, $G = G_Q \sum_n T_n$ ($G_S = 2G_Q\sum_n R^A_n$) and 
$F = \sum_n T_nR_n / \sum_n T_n$ 
($F_S = 2\sum_n R^A_n(1-R^A_n) / \sum_n R^A_n$) are the conductance 
and the Fano factor of the normal (superconducting) junction, 
and $G_Q = e^2/\pi$. The distribution of probabilities of electron-hole 
pair creations is characterized by 
$F_p = \sum_k p_k(1-p_k)/\sum_k p_k$, in analogy with the Fano factor 
for the transmission eigenvalues. 
The Bell test is analyzed in Fig.~\ref{fig:BellTestTunnel} 
as a function of the junction conductance, the amplitude of applied 
voltage, and different Fano factors that are close to the tunnel limit. 
We find that small deviations from the Poissonian statistics, that is, 
the presence of open channels, helps in violating the generalized Bell 
inequality. We note, that the Bell test also depends on the 
details of the drive through the probabilities $p_k$ and
distribution $F_p$ of electron-hole pair creations, which we will discuss later. It is also interesting by how much the inequality is violated for a given set of parameters. This extent of the Bell  inequality violation is shown in Fig.~\ref{fig:BI-cos-tun-N-SN} for the
harmonic drive. As we see, small amplitudes $V_0$ are in general favorable. In practice, a finite temperature sets an upper limit on $V_0$ for the Bell inequality violation, which we will discuss in Fig.~\ref{fig:finite-temperature}.
As regards the efficiency of the spin filters, we find that imperfect spin polarization decreases the possibility to violate our generalized Bell inequality \cite{supplemental}: in the low conductance normal junction the spin polarization has to be larger than $\mathcal{P}>2^{-1/4}\approx 84\%$, similar to the standard Bell test.

%%%%%%%%%%%%%%%%%%%%%%%%%%%%%%%%%%%%%%%%%%%%%%%%%%%%%%%%%%%%%%%%%%%%%%%%
%%%%%%%%%%%%%%%%%%%%%%%%%%%%%%%%%%%%%%%%%%%%%%%%%%%%%%%%%%%%%%%%%%%%%%%%

%%%%%%%%%%%%%%%%%%%%%%%%%%%%%%%%%%%%%%%%%%%%%%%%%%%%%%%%%%%%%%%%%%%%%%%%
\begin{figure}[b]
  %\centering
  \includegraphics[scale=1]{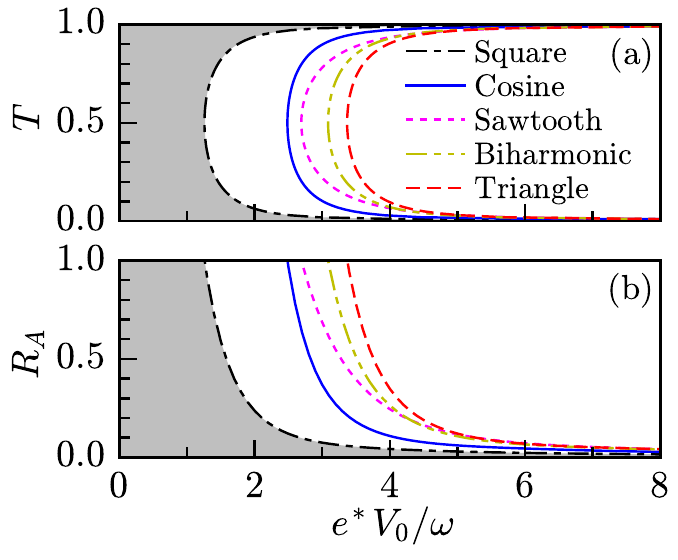}
  \caption{\label{fig:BellTestT1}Generalized Bell test for 
    a single-channel contact: (a) normal junction and (b) \textit{SN} beam 
    splitter. Results are shown for different shapes of the drive with 
    the amplitude $V_0$. 
    The biharmonic voltage 
    $V(t) = V_0[0.8 \cos(\omega t) + 0.1 \cos(3\omega t)]$ 
    is an approximation of the triangle-wave drive.
    The regions to the left of the displayed lines are where the 
    generalized Bell inequality is violated.}
\end{figure}
%%%%%%%%%%%%%%%%%%%%%%%%%%%%%%%%%%%%%%%%%%%%%%%%%%%%%%%%%%%%%%%%%%%%%%%%

The inequality Eq.~\eqref{eq:BI-final} can also be violated beyond 
the low-conductance limit 
in a quantum point contact with only a few 
discrete transport channels $T_n$. In the case of a single-channel contact, 
the parameters $X_i$ in Eq.~\eqref{eq:BI-final} read 
$X_1 = TR (\sum_k p_k + F_p - 1)$ for the normal junction and 
$X_2 = (R_A/4) (\sum_k p_k + F_p - 1)$ for the \textit{SN} beam splitter. 
The generalized Bell test is shown in Fig.~\ref{fig:BellTestT1} 
as a function of the contact transparency, amplitude of the drive, and
for different shapes of the drive voltage. In the normal junction, 
violation of Eq.~\eqref{eq:BI-final} occurs mostly in contacts 
that are either open or closed. This is because the probability of a 
successful Bell test is proportional to $TR$ which corresponds to one 
particle from a pair being transmitted to Bob while the 
other one is reflected to Alice, or vice versa. If this probability is
sufficiently low, the signal from one entangled pair is not obscured 
by other pairs and Eq.~\eqref{eq:BI-final} is violated.   
Similarly, the Bell test in the \textit{SN} beam splitter geometry is realized by the
injection of entangled pairs of electrons or holes with probability 
$R_A$ towards Alice and Bob. A violation of 
Eq.~\eqref{eq:BI-final} occurs mainly in \textit{SN} contacts with small $R_A$, 
in which the signal from one entangled pair is singled out from the
rest. As regards the shape of the drive, we note that the square-wave
drive creates more electron-hole pairs per cycle than the harmonic, 
sawtooth, or triangle-wave drive of the same amplitude, which decreases
the chances of entangled pair detection. In the single-channel contact
with low ac voltage, Eq.~\eqref{eq:BI-final} is violated for any contact
transparency. This is expected because in this case there is at most one
electron-hole pair created and detected per voltage cycle,
see Fig.~\ref{fig:BellTestTunnel}.

Finally, we would like to come back to the effect of a finite temperature $T_e$, since one expects that the entanglement detection will become more difficult if not impossible for increased temperatures. To this end, we have evaluated the correlators in Eq.~(\ref{BI-cumulant}) for a normal junction in the tunnel limit $\{T_n\}\ll 1$ and at arbitrary temperature \cite{supplemental}. The inequality is cast in the form $(\sqrt{2}-1)/4\leq \beta_{T_e}$, where we defined a new, temperature-dependent factor $\beta_{T_e}^2=\frac{G}{2G_Q}\sum_l |\cg_l|^2 l\coth(\frac{l\omega}{2T_e})$ with $\cg_l$ being the Fourier coefficients of the phase $e^{i\int_0^teV(t')dt'}$. The results are shown in Fig.~(\ref{fig:finite-temperature}). In the panel (a) an overview of the regions of entanglement detection in the conductance-voltage drive plot is given for different temperatures. An increasing temperature makes it more difficult to violate the inequality. For a given conductance, this leads to a prediction of a critical temperature of entanglement \cite{Beenakker:2005vx}, above which detection of entanglement is not possible. However, the numerical values still allow a violation for an experimentally accessible temperature range $T_e\lesssim \omega$.

%%%%%%%%%%%%%%%%%%%%%%%%%%%%%%%%%%%%%%%%%%%%%%%%%%%%%%%%%%%%%%%%%%%%%%%%
\begin{figure}[t]
	%\centering
	\includegraphics[scale=0.68]{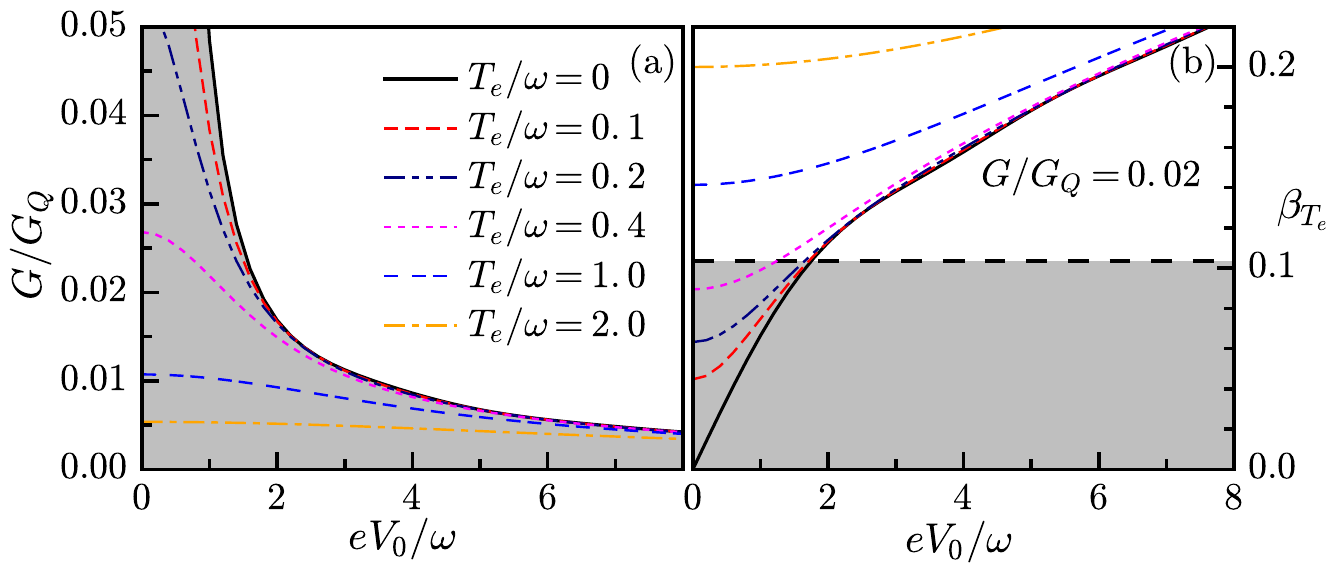}
	\caption{\label{fig:finite-temperature} Effect of a finite temperature on the violation of the generalized Bell inequality for a harmonic drive. (a) The voltage ranges of violation (to the left of the respective curves) are reduced by a finite temperature. (b) For a fixed conductance a violation is only possible below a certain critical temperature for entanglement generation.}
\end{figure}
%%%%%%%%%%%%%%%%%%%%%%%%%%%%%%%%%%%%%%%%%%%%%%%%%%%%%%%%%%%%%%%%%%%%%%%%

%%%%%%%%%%%%%%%%%%%%%%%%%%%%%%%%%%%%%%%%%%%%%%%%%%%%%%%%%%%%%%%%%%%%%%%%
%%%%%%%%%%%%%%%%%%%%%%%%%%%%%%%%%%%%%%%%%%%%%%%%%%%%%%%%%%%%%%%%%%%%%%%%
%\section{Summary}
%%%%%%%%%%%%%%%%%%%%%%%%%%%%%%%%%%%%%%%%%%%%%%%%%%%%%%%%%%%%%%%%%%%%%%%%
%%%%%%%%%%%%%%%%%%%%%%%%%%%%%%%%%%%%%%%%%%%%%%%%%%%%%%%%%%%%%%%%%%%%%%%%

In conclusion, we have developed a continuous-variable entanglement 
test which does not require charge quantization, single-particle 
detection, or unidirectional particle flow and is therefore suitable 
for entanglement detection in ac driven systems. 
This is in stark contrast to previous proposals, which were unrealistic due to temperature restrictions in experiment. In contrast, an ac drive mixes electron 
states of different energies and leads to a complex entangled state consisting of electron- and holelike 
quasiparticle pairs \cite{vanevic_electron_2016}.
We have shown that the entanglement between electrons and holes in this
situation can be probed using a four-terminal normal-metal 
junction, while the \textit{SN} beam splitter can be 
used to reveal the entanglement of Cooper pairs emitted or absorbed by 
the superconductor. 
The success of entanglement detection relies on the ability to 
differentiate between the overall current fluctuations and the specific 
current correlations coming from the entangled pairs. 
This can be achieved in quantum contacts with low conductance or a small 
number of transport channels and with an ac drive which creates few 
electron-hole pairs per cycle. 
We have investigated the feasibility of the entanglement test for a cosine, 
square, sawtooth, and triangle drive.
The ac drive affects the entanglement test through the total number of 
electron-hole pairs and the distribution of the pair creation 
probabilities.  Finally, we have shown that our proposed entanglement test is robust against a finite temperature in the experimentally accessible regime. A promising platform for the observation might be driven cold atomic quantum point contacts in which single-particle or -spin detection seems feasible.\cite{Lebrat:2019,Krinner:2017} In the future, it might be interesting to connect our findings to efforts to quantify many-body entanglement in correlated electron systems, which are so far based mainly on experimentally inaccessible quantities.

%%%%%%%%%%%%%%%%%%%%%%%%%%%%%%%%%%%%%%%%%%%%%%%%%%%%%%%%%%%%%%%%%%%%%%%%
%\begin{acknowledgments}
We gratefully acknowledge the support from DFG through SFB 767 
and the Serbian Ministry of Science Project No. 171027.
%\end{acknowledgments}
%%%%%%%%%%%%%%%%%%%%%%%%%%%%%%%%%%%%%%%%%%%%%%%%%%%%%%%%%%%%%%%%%%%%%%%%

%\newpage

%\appendix

% The \nocite command causes all entries in a bibliography to be 
% printed out whether or not they are actually referenced in the text. 
% This is appropriate for the sample file to show the different styles 
% of references, but authors most likely will not want to use it.
% \nocite{*}

%%%%%%%%%%%%%%%%%%%%%%%%%%%%%%%%%%%%%%%%%%%%%%%%%%%%%%%%%%%%%%%%%%%%%%%%
%%%%%%%%%%%%%%%%%%%%%%%%%%%%%%%%%%%%%%%%%%%%%%%%%%%%%%%%%%%%%%%%%%%%%%%%
%%%%%%%%%%%%%%%%%%%%%%%%%%%%%%%%%%%%%%%%%%%%%%%%%%%%%%%%%%%%%%%%%%%%%%%%
%%%%%%%%%%%%%%%%%%%%%%%%%%%%%%%%%%%%%%%%%%%%%%%%%%%%%%%%%%%%%%%%%%%%%%%%
%\bibliographystyle{apsrev}
%\bibliography{references}

%
%%%%%%%%%%%%%%%%%%%%%%%%%%%%%%%%%%%%%%%%%%%%%%%%%%%%%%%%%%%%%%%%%%%%%%%%
%%%%%%%%%%%%%%%%%%%%%%%%%%%%%%%%%%%%%%%%%%%%%%%%%%%%%%%%%%%%%%%%%%%%%%%%
%%%%%%%%%%%%%%%%%%%%%%%%%%%%%%%%%%%%%%%%%%%%%%%%%%%%%%%%%%%%%%%%%%%%%%%%
%%%%%%%%%%%%%%%%%%%%%%%%%%%%%%%%%%%%%%%%%%%%%%%%%%%%%%%%%%%%%%%%%%%%%%%%
%\newpage

\clearpage

\section{Supplemental material for
	"Continuous-variable entanglement test in a driven quantum 
	contact"}
Here we present the derivation of Eq.~(3) of the main text for the cumulant generating functions $\cS_i$ of the normal junction and the superconductor -- normal-metal beam splitter and discuss the finite-temperature effect on the normal junction. In the end, we also discuss the Bell test for a single-channel normal-metal beam splitter.

%%%%%%%%%%%%%%%%%%%%%%%%%%%%%%%%%%%%%%%%%%%%%%%%%%%%%%%%%%%%%%%%%%%%%%%%
\section{Normal-metal junction}
%%%%%%%%%%%%%%%%%%%%%%%%%%%%%%%%%%%%%%%%%%%%%%%%%%%%%%%%%%%%%%%%%%%%%%%%
The junction is schematically depicted in Fig.~1(a) of the main text. There are 4 leads: 2 left leads for Alice and 2 right leads for Bob. We assume the leads are perfectly spin filtered in directions $\pm\mbf a$ and $\pm\mbf b$ where $|\mbf a|=|\mbf b|=1$. We count charges that enter each of the leads and assign the counting fields $\chi_j^\sigma$ ($j=A,B$; $\sigma=\uparrow,\downarrow$) to them. The leads are characterized by the Keldysh-Green's functions
%%%%%%%%%%%%%%%%%%%%%%%%%%%%%%%%%%%%%%%%%%%%%%%%%%%%%%%%%%%%%%%%%%%%%%%%
\begin{align}
\vG_L(\chi_A^\sigma) 
&= 
e^{-i\chi_A^\sigma \check\tau_1/2} \, \vG_L(0) \, e^{i\chi_A^\sigma 
	\check\tau_1/2},
\\
\vG_R(\chi_B^\sigma) 
&= 
e^{-i\chi_B^\sigma \check\tau_1/2} \, \vG_R(0) \, e^{i\chi_B^\sigma 
	\check\tau_1/2},
\end{align}
%%%%%%%%%%%%%%%%%%%%%%%%%%%%%%%%%%%%%%%%%%%%%%%%%%%%%%%%%%%%%%%%%%%%%%%%
where 
%%%%%%%%%%%%%%%%%%%%%%%%%%%%%%%%%%%%%%%%%%%%%%%%%%%%%%%%%%%%%%%%%%%%%%%%
\begin{equation}
\vG_l(0) = 
\begin{pmatrix}
1 & 2h_l \\
0 & -1
\end{pmatrix}.
\end{equation}
%%%%%%%%%%%%%%%%%%%%%%%%%%%%%%%%%%%%%%%%%%%%%%%%%%%%%%%%%%%%%%%%%%%%%%%%
Here, $\check \tau_1$ is the first Pauli matrix in Keldysh 
space and $f_l=(1-h_l)/2$ ($l=L,R$) are the generalized nonequilibrium 
distribution functions which depend on two time or energy 
indices. For an ac driven junction, we can use $2\times 2$ 
representation of $h_L$ and $h_R$ in energy space 
\cite{vanevic_elementary_2007,*vanevic_elementary_2008}
%%%%%%%%%%%%%%%%%%%%%%%%%%%%%%%%%%%%%%%%%%%%%%%%%%%%%%%%%%%%%%%%%%%%%%%%
\begin{equation}
h_L = 
\begin{pmatrix}
0 & e^{-i\alpha_k} \\
e^{i\alpha_k} & 0
\end{pmatrix},
\quad
h_R =
\begin{pmatrix}
0 & 1 \\
1 & 0
\end{pmatrix},
\end{equation}
%%%%%%%%%%%%%%%%%%%%%%%%%%%%%%%%%%%%%%%%%%%%%%%%%%%%%%%%%%%%%%%%%%%%%%%%
where $p_k = \sin^2(\alpha_k/2)$ ($k=1,2,\ldots$) are the probabilities 
of electron-hole pair creations. Note that the Green's functions 
$\vG_l$ are scalars in the spin space. 

For simplicity, we assume that the conductances of the spin filters are 
much larger than the one of the central junction 
$g=(e^2/\pi)\sum_n T_n$. In this case, the electron which arrives 
at Alice or Bob will enter the spin-polarized leads without 
backscattering. Since $A$ is strongly coupled to the spin-filtering 
leads and only weakly coupled to $B$, we can obtain the Green's 
function $\vG_A$ of the node $A$ with the node $B$ disconnected. 
In addition, we assume that the spin filters have the same conductance 
and that an electron can enter any of the spin-filtering leads with 
equal probability. The Green's function $\vG_A$ is obtained from the 
matrix current conservation
$[ (1+\mbf a\cdot \hat{\mbf\sigma})\vG_L(\chi_A^\uparrow) 
+ (1-\mbf a\cdot \hat{\mbf\sigma})\vG_L(\chi_A^\downarrow), \vG_A] = 0$ 
and the normalization condition $\vG_A^2=\check 1$ 
\cite{lorenzo_full_2005}. 
Here, $\hat{\mbf \sigma}$ is the vector of Pauli matrices in spin space.
We obtain 
%%%%%%%%%%%%%%%%%%%%%%%%%%%%%%%%%%%%%%%%%%%%%%%%%%%%%%%%%%%%%%%%%%%%%%%%
\begin{equation}\label{GA}
\vG_A(\chi_A^\uparrow,\chi_A^\downarrow)
=
\frac{1+\mbf a\cdot\hat{\mbf \sigma}}{2}\vG_L(\chi_A^\uparrow) +
\frac{1-\mbf a\cdot\hat{\mbf \sigma}}{2}\vG_L(\chi_A^\downarrow),
\end{equation}
%%%%%%%%%%%%%%%%%%%%%%%%%%%%%%%%%%%%%%%%%%%%%%%%%%%%%%%%%%%%%%%%%%%%%%%%
and similarly for the node $B$, 
%%%%%%%%%%%%%%%%%%%%%%%%%%%%%%%%%%%%%%%%%%%%%%%%%%%%%%%%%%%%%%%%%%%%%%%%
\begin{equation}\label{GB}
\vG_B(\chi_B^\uparrow,\chi_B^\downarrow)
=
\frac{1+\mbf b\cdot\hat{\mbf \sigma}}{2}\vG_R(\chi_B^\uparrow) +
\frac{1-\mbf b\cdot\hat{\mbf \sigma}}{2}\vG_R(\chi_B^\downarrow).
\end{equation}
%%%%%%%%%%%%%%%%%%%%%%%%%%%%%%%%%%%%%%%%%%%%%%%%%%%%%%%%%%%%%%%%%%%%%%%%

Cumulant generating function for the normal-metal junction is given by 
%%%%%%%%%%%%%%%%%%%%%%%%%%%%%%%%%%%%%%%%%%%%%%%%%%%%%%%%%%%%%%%%%%%%%%%%
\begin{equation}\label{CGF1}
\cS_1(\chi) = \frac{1}{2} \sum_n {\rm Tr}\ln
\left(
\check 1 + \frac{T_n}{4}(\aCom{\vG_A}{\vG_B}-2) 
\right),
\end{equation}
%%%%%%%%%%%%%%%%%%%%%%%%%%%%%%%%%%%%%%%%%%%%%%%%%%%%%%%%%%%%%%%%%%%%%%%%
where the trace is taken over Keldysh, spin, and energy indices. 
Since Alice and Bob are measuring the differences between the spin-up 
and spin-down charges, $\hat A \equiv \hat N_A^\uparrow - \hat 
N_A^\downarrow = \int dt(\hat I_A^\uparrow - \hat I_A^\downarrow)/e$
and $\hat B \equiv \hat N_B^\uparrow - \hat N_B^\downarrow
=\int dt (\hat I_B^\uparrow - \hat I_B^\downarrow)/e$, we can set
$\chi_A^\uparrow = \chi_A$, $\chi_A^\downarrow = -\chi_A$,  
$\chi_B^\uparrow = \chi_B$, and $\chi_B^\downarrow = -\chi_B$, where 
$\chi_A$ and $\chi_B$ are the counting fields related to the statistics 
of $\hat A$ and $\hat B$. 

To compute $\cS_1$, we proceed as follows. We note that $\vG_A$ and
$\vG_B$ are $8\times 8$ matrices in Keldysh $\times$ spin $\times$ 
energy space. Cumulant generating function $\cS_1$ can be written as a 
sum $\cS_1 = \cS_1^+ + \cS_1^-$, where
%%%%%%%%%%%%%%%%%%%%%%%%%%%%%%%%%%%%%%%%%%%%%%%%%%%%%%%%%%%%%%%%%%%%%%%%
\begin{equation}
\cS_1^\pm (\chi_A,\chi_B) =
\frac{1}{2} \sum_n {\rm Tr}\ln
\left(
\check 1 \pm \frac{\sqrt{T_n}}{2}(\vG_A - \vG_B)
\right).
\end{equation}
%%%%%%%%%%%%%%%%%%%%%%%%%%%%%%%%%%%%%%%%%%%%%%%%%%%%%%%%%%%%%%%%%%%%%%%%
It turns out that $\check 1 \pm (\sqrt{T_n}/2)(\vG_A - \vG_B)$ 
have the same eigenvalues, hence $\cS_1^+ = \cS_1^-$. Moreover, 
the eigenvalues are pairwise equal to each other, 
$\lambda_1=\lambda_2$, $\lambda_3=\lambda_4$, $\lambda_5=\lambda_6$, and
$\lambda_7=\lambda_8$. We obtain that the product  
$\lambda_1\lambda_3\lambda_5\lambda_7 = 1 + 2p_kT_nR_n
[\cos(\chi_A)\cos(\chi_B)-1 + \mbf a\cdot\mbf b 
\sin(\chi_A)\sin(\chi_B)]$. Therefore, the cumulant generating function 
reads
%%%%%%%%%%%%%%%%%%%%%%%%%%%%%%%%%%%%%%%%%%%%%%%%%%%%%%%%%%%%%%%%%%%%%%%%
\begin{multline}
\cS_1(\chi_A,\chi_B) = \frac{\tau\omega}{\pi} \sum_{n,k}
\ln \Big( 
1 + 2p_kT_nR_n \\
\times [\cos(\chi_A)\cos(\chi_B) - 1
+ \mbf a\cdot \mbf b \sin(\chi_A)\sin(\chi_B)]
\Big).
\end{multline}
%%%%%%%%%%%%%%%%%%%%%%%%%%%%%%%%%%%%%%%%%%%%%%%%%%%%%%%%%%%%%%%%%%%%%%%%
This expression reduces to Eq. (3) of the main text. 

%%%%%%%%%%%%%%%%%%%%%%%%%%%%%%%%%%%%%%%%%%%%%%%%%%%%%%%%%%%%%%%%%%%%%%%%
\section{Superconductor -- normal-metal beam splitter}
%%%%%%%%%%%%%%%%%%%%%%%%%%%%%%%%%%%%%%%%%%%%%%%%%%%%%%%%%%%%%%%%%%%%%%%%

The junction consists of a superconductor ($S$) coupled to the 
normal-metal lead ($N$) through the central junction characterized by 
transmission eigenvalues $\{T_n\}$. The normal lead is split into 4 
outgoing terminals with spin filters along directions $\pm\mbf a$ 
(Alice) and $\pm\mbf b$ (Bob), see Fig.~1(b). As before, we assume 
strong coupling of the spin filters to the node $N$ and neglect the 
backscattering. We also assume that the spin-filtering leads have the 
same conductance. The cumulant generating function is given by
%%%%%%%%%%%%%%%%%%%%%%%%%%%%%%%%%%%%%%%%%%%%%%%%%%%%%%%%%%%%%%%%%%%%%%%%
\begin{equation}\label{eqSup:CGF}
\cS_2(\chi) = \frac{1}{4} \sum_n {\rm Tr} \ln
\left(
1 + \frac{T_n}{4} (\aCom{\mbf G_S(0)}{\mbf G_N(\chi)}-2)
\right),
\end{equation}
%%%%%%%%%%%%%%%%%%%%%%%%%%%%%%%%%%%%%%%%%%%%%%%%%%%%%%%%%%%%%%%%%%%%%%%%
where the trace is taken in electron-hole (Nambu), Keldysh, spin, and
energy indices. 

At energies and temperatures much smaller than the gap, the Green's
function $\mbf G_S$ of the superconductor is given by
%%%%%%%%%%%%%%%%%%%%%%%%%%%%%%%%%%%%%%%%%%%%%%%%%%%%%%%%%%%%%%%%%%%%%%%%
\begin{equation}
\mbf G_S(0) = \bar\tau_2\otimes \check 1 =
\begin{pmatrix}
0 & -i\check 1 \\
i \check 1 & 0
\end{pmatrix},
\end{equation}
%%%%%%%%%%%%%%%%%%%%%%%%%%%%%%%%%%%%%%%%%%%%%%%%%%%%%%%%%%%%%%%%%%%%%%%%
where the block-matrix structure is in electron-hole space. 
The Green's function $\mbf G_N(\chi)$ of the node $N$ in the normal
terminal is given by the matrix current conservation
%%%%%%%%%%%%%%%%%%%%%%%%%%%%%%%%%%%%%%%%%%%%%%%%%%%%%%%%%%%%%%%%%%%%%%%%
\begin{equation}\label{eqSup:GNcom}
\left[
\begin{pmatrix}
\vG_{Ae} + \vG_{Be} & 0 \\
0 & -(\vG_{Ah} + \vG_{Bh})
\end{pmatrix}, \mbf G_N
\right] = 0
\end{equation}
%%%%%%%%%%%%%%%%%%%%%%%%%%%%%%%%%%%%%%%%%%%%%%%%%%%%%%%%%%%%%%%%%%%%%%%%
and the normalization condition $\mbf G_N^2 = 1$. 
Here, the Green's functions for electrons and holes are given by
%%%%%%%%%%%%%%%%%%%%%%%%%%%%%%%%%%%%%%%%%%%%%%%%%%%%%%%%%%%%%%%%%%%%%%%%
\begin{align}
\vG_{Ae} & = 
\frac{1+\mbf a\cdot\hat{\mbf\sigma}}{2} \vG_e(\chi_A^\uparrow)
+ \frac{1-\mbf a\cdot\hat{\mbf\sigma}}{2} \vG_e(\chi_A^\downarrow), \\
\vG_{Be} & = 
\frac{1+\mbf b\cdot\hat{\mbf\sigma}}{2} \vG_e(\chi_B^\uparrow)
+ \frac{1-\mbf b\cdot\hat{\mbf\sigma}}{2} \vG_e(\chi_B^\downarrow), 
\end{align}
%%%%%%%%%%%%%%%%%%%%%%%%%%%%%%%%%%%%%%%%%%%%%%%%%%%%%%%%%%%%%%%%%%%%%%%%
and
%%%%%%%%%%%%%%%%%%%%%%%%%%%%%%%%%%%%%%%%%%%%%%%%%%%%%%%%%%%%%%%%%%%%%%%%
\begin{align}
\vG_{Ah} & = 
\frac{1-\mbf a\cdot\hat{\mbf\sigma}}{2} \vG_h(-\chi_A^\uparrow)
+ \frac{1+\mbf a\cdot\hat{\mbf\sigma}}{2} \vG_h(-\chi_A^\downarrow), \\
\vG_{Bh} & = 
\frac{1-\mbf b\cdot\hat{\mbf\sigma}}{2} \vG_h(-\chi_B^\uparrow)
+ \frac{1+\mbf b\cdot\hat{\mbf\sigma}}{2} \vG_h(-\chi_B^\downarrow).
\end{align}
%%%%%%%%%%%%%%%%%%%%%%%%%%%%%%%%%%%%%%%%%%%%%%%%%%%%%%%%%%%%%%%%%%%%%%%%
We note that the counting fields and spin-polarization vectors for 
holes are the opposite from the ones for electrons. The counting fields 
are incorporated via the gauge transform
$\vG_{e,h}(\chi) = e^{-i \chi\bar \tau_3\otimes \check \tau_1 /2} \,
\vG_{e,h}(0) \, e^{i \chi\bar \tau_3\otimes \check \tau_1 /2}$,
with
%%%%%%%%%%%%%%%%%%%%%%%%%%%%%%%%%%%%%%%%%%%%%%%%%%%%%%%%%%%%%%%%%%%%%%%%
\begin{equation}
\vG_e(0) = 
\begin{pmatrix}
1 & 2UhU^\dagger \\
0 & -1
\end{pmatrix},
\quad
\vG_h(0) = 
\begin{pmatrix}
1 & 2U^\dagger h U \\
0 & -1
\end{pmatrix}.
\end{equation}
%%%%%%%%%%%%%%%%%%%%%%%%%%%%%%%%%%%%%%%%%%%%%%%%%%%%%%%%%%%%%%%%%%%%%%%%
Here, $U(t',t'')=\exp[-i\int_0^{t'} eV(t)dt]\delta(t'-t'')$  
is unitary operator which takes into account the time-dependent drive
\cite{vanevic_elementary_2007,*vanevic_elementary_2008}. 

From Eq.~\eqref{eqSup:GNcom} we obtain 
%%%%%%%%%%%%%%%%%%%%%%%%%%%%%%%%%%%%%%%%%%%%%%%%%%%%%%%%%%%%%%%%%%%%%%%%
\begin{equation}
\mbf G_N = 
\begin{pmatrix}
\vG_{Ne} & 0 \\
0 & -\vG_{Nh}
\end{pmatrix},
\quad
\vG_{Ne,h} = \frac{\vG_{Ae,h}+\vG_{Be,h}}{2}.
\end{equation}
%%%%%%%%%%%%%%%%%%%%%%%%%%%%%%%%%%%%%%%%%%%%%%%%%%%%%%%%%%%%%%%%%%%%%%%%
Substituting this result in Eq.~\eqref{eqSup:CGF} and after taking the 
trace over electron-hole indices we find
%%%%%%%%%%%%%%%%%%%%%%%%%%%%%%%%%%%%%%%%%%%%%%%%%%%%%%%%%%%%%%%%%%%%%%%%
\begin{equation}\label{eqSup:CGF1}
\cS_2 = \frac{1}{2} \sum_n {\rm Tr}\ln
\left(
\check 1 + \frac{\sqrt{R^A_n}}{2}
\Big( \vG_{Ne}(2V(t)) - \vG_{Nh}(0) \Big)
\right).
\end{equation}
%%%%%%%%%%%%%%%%%%%%%%%%%%%%%%%%%%%%%%%%%%%%%%%%%%%%%%%%%%%%%%%%%%%%%%%%
Here, the remaining trace is taken over Keldysh, spin, and energy 
indices and $R^A_n = T_n^2/(2-T_n)^2$ are the coefficients of Andreev 
reflections. In Eq.~\eqref{eqSup:CGF1} we have also performed a gauge 
transform under the trace and ascribed the effect of the voltage 
drive to the electron part of the Green's function, which leads to
an effective voltage (or charge) doubling 
\cite{vanevic_elementary_2016}. 
After diagonalization of the operator in Eq.~\eqref{eqSup:CGF1}, 
we obtain
%%%%%%%%%%%%%%%%%%%%%%%%%%%%%%%%%%%%%%%%%%%%%%%%%%%%%%%%%%%%%%%%%%%%%%%%
\begin{multline}
\cS_2(\chi_A,\chi_B) = \frac{\tau\omega}{\pi} \sum_{n,k} \ln
\bigg(
1 + \frac{p_kR^A_n}{4} \\
\times \sum_{\alpha,\beta = \pm 1}
\frac{1+\alpha\beta\mbf a \cdot \mbf b}{2}
(e^{i\alpha\chi_A - i\beta\chi_B}-1)
\bigg).
\end{multline}
%%%%%%%%%%%%%%%%%%%%%%%%%%%%%%%%%%%%%%%%%%%%%%%%%%%%%%%%%%%%%%%%%%%%%%%%
This expression coincides with Eq.~(3) of the main text. 

%%%%%%%%%%%%%%%%%%%%%%%%%%%%%%%%%%%%%%%%%%%%%%%%%%%%%%%%%%%%%%%%%%%%%%%%
\section{Finite temperature effect on normal-metal junction}
%%%%%%%%%%%%%%%%%%%%%%%%%%%%%%%%%%%%%%%%%%%%%%%%%%%%%%%%%%%%%%%%%%%%%%%%
At finite temperature, $h_L$ and $h_R$ cannot be written as $2\times2$
matrices because the involution property does not hold: $h^2_L\neq h^2_R\neq 1$.
In energy representation, they are written as \cite{vanevic_elementary_2008}
\begin{equation}\label{hl}
h_L(\epsilon',\epsilon'')=\sum_{l,m}\cg_l\cg^*_{l+m}h(\epsilon'-l\omega-e\bar{V})2\pi\delta(\epsilon''-\epsilon'-m\omega)
\end{equation}
\begin{equation}\label{hr}
h_R(\epsilon',\epsilon'')=h(\epsilon')2\pi\delta(\epsilon'-\epsilon'')
\end{equation}
where
\begin{equation}\label{hh}
h(\epsilon)=\tanh\left({\frac{\epsilon}{2T_e}}\right) 
\end{equation}
and $\cg_l$ is the $l$th coefficient of the Fourier series of the ac phase $e^{i\phi(t)}=\sum_l\cg_le^{-il\omega t}$ and $\phi(t)=\int^t_0e\Delta V(t')dt'$.
In tunnel limit, the cumulant generating function Eq.~\eqref{CGF1} becomes
\begin{equation}
\cS_1(\chi)\approx\frac{1}{4}\frac{G}{G_Q}Tr_{\epsilon}(\vG_A\vG_B)
\end{equation}
where the trace $Tr_{\epsilon}$ is taken in energy indices. After substituting Eqs.~\eqref{hl}-\eqref{hh} into $\cS_1$ and integration over energy, we have
\begin{align}\label{cS1_1}
\cS_1(\chi)=&\frac{G}{G_Q}\sum_l|\cg_l|^2l\coth(\frac{l\omega}{2T_e}) \nonumber \\
&\times[\cos(\chi_A)\cos(\chi_B)+\mbf a\cdot \mbf b\sin(\chi_A)\sin(\chi_B)-1]
\end{align}
Here we have assumed zero dc bias. Eq.~\eqref{cS1_1} could be equivalently written as
\begin{align}\label{cS1_2}
\cS_1(\chi)=&\frac{G}{2G_Q}\sum_l|\cg_l|^2 l\coth(\frac{l\omega}{2T_e}) \nonumber \\
&\times\sum_{\alpha,\beta=\pm 1}\frac{1+\alpha\beta \mbf a\cdot \mbf b}{2}\left(e^{i\alpha\chi_A-i\beta\chi_B}-1\right)
\end{align}
Taking derivatives of Eq.~\eqref{cS1_2} with respect to the counting fields, we have the relations between the cumulants
\begin{align}
\llangle A^{2n}\rrangle=&\llangle B^{2n}\rrangle=\llangle A^{2n}B^{2n}\rrangle =\frac{G}{G_Q}\sum_l |\cg_l|^2 l\coth(\frac{l\omega}{2T_e}) \\
\llangle AB\rrangle=&\llangle A^3B\rrangle=\llangle AB^3\rrangle=-\mbf a\cdot \mbf b \llangle A^{2n}\rrangle
\end{align}
Choosing the angles between spin-polarization directions $\theta_{\mbf a\mbf b}=\theta_{\mbf a'\mbf b}=\theta_{\mbf a\mbf b'}=\pi/4$ and $\theta_{\mbf a'\mbf b'}=3\pi/4$, the generalized Bell inequality (Eq.~(4) of the main text) reduces to
\begin{equation}\label{BI}
\frac{2-\sqrt{2}}{4} \leq \frac{\langle A^2\rangle}{\sqrt{\langle A^4\rangle}}
\end{equation}
Let $X_{T_e}=\frac{G}{2G_Q}\sum_l |\cg_l|^2 l\coth(\frac{l\omega}{2T_e})$, Eq.~\eqref{BI} can be rewritten as
\begin{equation}\label{BI1}
\frac{\sqrt{2}-1}{4} \leq \sqrt{\frac{X_{T_e}}{1+6X_{T_e}}}
\end{equation}
which takes the same form of Eq.~(2) of the main text. In tunnel limit, $G/G_Q \ll 1$, thus Eq.~\eqref{BI1} can be approximately written as $(\sqrt{2}-1)/4\leq\beta_{T_e}$ where $\beta_{T_e}^2=X_{T_e}$.

%%%%%%%%%%%%%%%%%%%%%%%%%%%%%%%%%%%%%%%%%%%%%%%%%%%%%%%%%%%%%%%%%%%%%%%%
\section{Bell test for a single-channel normal-metal beam splitter}
\begin{figure}[t]
	%\centering
	\includegraphics[scale=1.]{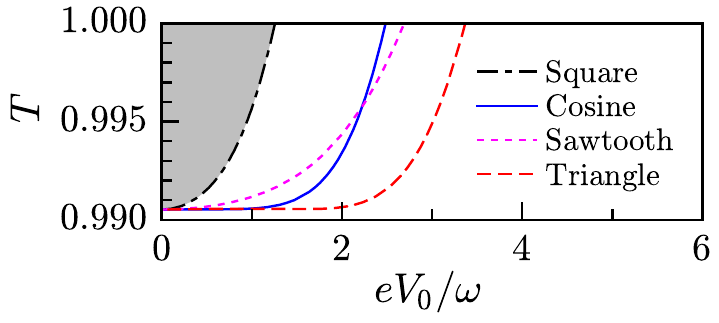}
	\caption{\label{fig:NN-beam-splitter} Generalized Bell test for 
		a single-channel normal-metal beam splitter. Results are shown for different shapes of the drive with the amplitude $V_0$. 
		The generalized Bell inequality is 
		violated in the regions above the displayed lines where contact 
		transparencies are very close to $T=1$.}
\end{figure}

%%%%%%%%%%%%%%%%%%%%%%%%%%%%%%%%%%%%%%%%%%%%%%%%%%%%%%%%%%%%%%%%%%%%%%%%
In the following we analyze a normal-metal beam splitter 
analogous to the superconducting one shown in 
Fig.~1(b) of the main text. It has been found in 
\cite{chtchelkatchev_bell_2002} that the normal beam splitter  
is not suitable for detection of entangled electron spin singlets 
injected 
% into the system 
by a dc voltage \cite{lorenzo_full_2005}. We show that this 
setup is inefficient in a driven case as well. For the cumulant 
generating function we find 
%%%%%%%%%%%%%%%%%%%%%%%%%%%%%%%%%%%%%%%%%%%%%%%%%%%%%%%%%%%%%%%%%%%%%%%%
$\cS_N(\chi_A,\chi_B) = M\sum_{n,k}\ln 
[1+(p_kT_nR_n/2)(e^{i\chi_A}+e^{-i\chi_A}+e^{i\chi_B}+e^{-i\chi_B}-4)
+(p_kT_n^2/8)\sum_{\alpha,\beta=\pm 1}(1+\alpha\beta\mbf a\cdot \mbf 
b)(e^{i\alpha\chi_A-i\beta\chi_B}-1)]$. 
%%%%%%%%%%%%%%%%%%%%%%%%%%%%%%%%%%%%%%%%%%%%%%%%%%%%%%%%%%%%%%%%%%%%%%%%
The charge transfer statistics consists of two parts. 
The spin-dependent part is proportional to $p_kT_n^2$ and is related to 
the Bell-type events in which both electron and hole from a pair are 
transmitted to Alice and Bob. The spin-independent 
part is proportional to $p_kT_nR_n$ and describes events in 
which only one particle is transmitted to Alice or Bob while the other 
one is reflected back at the contact. These events carry no information 
on the particle spin correlations and lead to current fluctuations 
which obscure entanglement detection. Therefore, a violation of a 
generalized Bell inequality can be achieved only in a normal beam 
splitter of large transparency. For a single-channel junction with 
transmission coefficient $T$, Eq.~(4) of the main text reduces to 
$1-1/\sqrt{2} \le 2R/(1+3X_p/2) + 2\sqrt{(2R+X_p/2)/(1+3X_p/2)}$, 
where $X_p = \sum_k p_k + F_p - 1$. This inequality is violated
for $T>0.99$ when Bell-type events dominate over particle reflections 
at the contact, see Fig.~\ref{fig:NN-beam-splitter}.  

%%%%%%%%%%%%%%%%%%%%%%%%%%%%%%%%%%%%%%%%%%%%%%%%%%%%%%%%%%%%%%%%%%%%%%%%
%%%%%%%%%%%%%%%%%%%%%%%%%%%%%%%%%%%%%%%%%%%%%%%%%%%%%%%%%%%%%%%%%%%%%%%%

%%%%%%%%%%%%%%%%%%%%%%%%%%%%%%%%%%%%%%%%%%%%%%%%%%%%%%%%%%%%%%%%%%%%%%%%
\section{Bell test for normal-metal junction with finite-polarized spin filters}
\begin{figure}[t]
	%\centering
	\includegraphics[scale=1.]{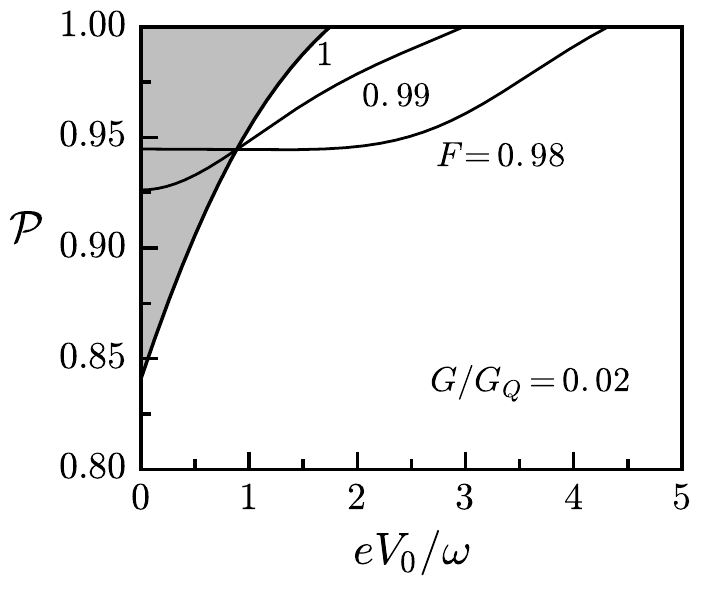}
	\caption{\label{fig:finite-polarization} Generalized Bell test for 
		a normal junction with finite polarization for harmonic drive. Fano factors $F=1,0.99,0.98$ are shown in the plot. The generalized Bell inequality is violated for spin polarizations and applied voltages that are above the lines.}
\end{figure}

%%%%%%%%%%%%%%%%%%%%%%%%%%%%%%%%%%%%%%%%%%%%%%%%%%%%%%%%%%%%%%%%%%%%%%%%
In this section we consider the normal junction where the spin filters 
are not fully-polarized, namely $\mbf a = \mathcal{P}\mbf{m_a}$ and
$\mbf b = \mathcal{P}\mbf{m_b}$ where spin polarization 
$0\le \mathcal{P}\le 1$ and $|\mbf{m_a}|=|\mbf{m_b}|=1$.
The cumulant generating function is given by Eqs.~\eqref{GA} - 
\eqref{CGF1} where $\chi_A^\uparrow = -\chi_A^\downarrow = \chi_A$
and $\chi_B^\uparrow = -\chi_B^\downarrow = \chi_B$.
For the cumulant generating function we find
\begin{multline}
\cS_1(\chi)=\sum_{n,k}\ln\Big[1+2T_n(1-T_n)p_k \\
\times[\cos(\chi_A)\cos(\chi_B)-1+\mbf{m_a}\cdot\mbf{m_b}\mathcal{P}^2\sin(\chi_A)\sin(\chi_B)]
\\
+\frac{T_n^2}{2}p_k(1-\mathcal{P}^2)[\cos(2\chi_A)+\cos(2\chi_B)-2] 
+\frac{T_n^2}{4}p_k^2(1-\mathcal{P}^2)^2 \\
\times [1-\cos(2\chi_A)-\cos(2\chi_B)+\cos(2\chi_A)\cos(2\chi_B)]
\Big].
\end{multline}
Note that for $\mathcal{P}\ne 1$ there are no simple relations between 
the cumulants. In the low conductance limit and by choosing the angles 
between spin-polarization directions $\theta_{\mbf a\mbf 
	b}=\theta_{\mbf a'\mbf b}=\theta_{\mbf a\mbf b'}=\pi/4$ and 
$\theta_{\mbf a'\mbf b'}=3\pi/4$, the generalized Bell inequality reads
\begin{equation}\label{BI-finite-polarization}
\frac{\sqrt{2}\mathcal{P}^2-1}{4}\leq \beta_{\mathcal{P}}
\end{equation}
where
\begin{align}
\beta_{\mathcal{P}}^2=&(2-\mathcal{P}^4)\frac{G}{G_Q}\sum_k p_k 
-(3-2\mathcal{P}^2)(1-F_p)(1-F) \nonumber \\
&+2(1-\mathcal{P}^2)(1-F).
\end{align}
In the fully-polarized spin case $\mathcal{P}=1$, the above inequality 
reduces to Eq.~(5) of the main text. The violations of the above 
inequality for different Fano factors are plotted in 
Fig.~\ref{fig:finite-polarization}. 
%%%%%%%%%%%%%%%%%%%%%%%%%%%%%%%%%%%%%%%%%%%%%%%%%%%%%%%%%%%%%%%%%%%%%%%%

\end{document}